# Lorenz or Coulomb in Galilean Electromagnetism ?

**Germain Rousseaux**
INLN – UMR 6618 CNRS,
1361 route des Lucioles,
06560 Valbonne

**Résumé.** L'électromagnetisme galiléen fut découvert il y a trente ans par Lévy-Leblond & Le Bellac. Cependant, ces auteurs ont uniquement exploré les consequences pour les champs et non pour les potentiels. En suivant De Montigny & al., nous montrons que la condition de jauge de Coulomb est la limite magnétique de la condition de jauge de Lorenz alors que cette dernière s'applique dans la limite électrique de Lévy-Leblond & Le Bellac. Contrairement à De Montigny & al., nous utilisons des ordres de grandeurs motivés par des considérations physiques dans notre démonstration.

**Abstract.** Galilean Electromagnetism was discovered thirty years ago by Levy-Leblond & Le Bellac. However, these authors only explored the consequences for the fields and not for the potentials. Following De Montigny & al., we show that the Coulomb gauge condition is the magnetic limit of the Lorenz gauge condition whereas the Lorenz gauge condition applies in the electric limit of Lévy-Leblond & Le Bellac. Contrary to De Montigny & al. who used Galilean tensor calculus, we use orders of magnitude based on physical motivations in our derivation.

## 1. Introduction

Does there exist a galilean limit of Maxwell equations ? According to Lévy-Leblond & Le Bellac, the answer is positive [1]. Indeed, they have shown that there exist not one as in mechanics but two well defined Galilean limits of the full set of Maxwell equations : the magnetic limit and the electric limit. The two Galilean limits were introduced by Lévy-Leblond and Le Bellac without demonstration. More precisely, they have show that two particular approximations of the full set of Maxwell equations were compatible with the two Galilean transformations for the field they "derived".

If one denotes $\gamma = \left(1 - \mathbf{v}^2/c_L^2\right)^{-1/2}$ where $c_L$ is the light velocity, the relativistic transformations for the fields in vacuum between two inertial frames with relative velocity $\mathbf{v}$ are :

$$\mathbf{E}' = \gamma(\mathbf{E} + \mathbf{v} \times \mathbf{B}) + \frac{(1-\gamma)(\mathbf{v}.\mathbf{E})\mathbf{v}}{\mathbf{v}^2} \quad \text{and} \quad \mathbf{B}' = \gamma\left(\mathbf{B} - \left(1/c_L^2\right)\mathbf{v} \times \mathbf{E}\right) + \frac{(1-\gamma)(\mathbf{v}.\mathbf{B})\mathbf{v}}{\mathbf{v}^2}$$

In vacuum, one obtains the magnetic limit by stating that $|\mathbf{v}|/c_L \ll 1$ and $E \ll c_L B$. Conversely, the electric limit is obtained by stating that $|\mathbf{v}|/c_L \ll 1$ and $E \gg c_L B$. Hence, one ends up with two set of low-velocity formula from the Lorentz transformations [1]:

<u>Electric Limit :</u>
$$\mathbf{E}' = \mathbf{E} \quad \text{and} \quad \mathbf{B}' = \mathbf{B} - \left(1/c_L^2\right)\mathbf{v} \times \mathbf{E}$$

<u>Magnetic Limit :</u>
$$\mathbf{E}' = \mathbf{E} + \mathbf{v} \times \mathbf{B} \quad \text{and} \quad \mathbf{B}' = \mathbf{B}$$

These two limits are practically very important since they correspond to the so-called electroquasistatic and magnetoquasistatic approximations of engineering electromagnetism as described in Haus & Melcher e-book [2]. Moreover, magnetohydrodynamics relies on the magnetic limit whereas electrohydrodynamics relies on the electric limit of Maxwell equations.

Severals authors have discussed recently Lévy-Leblond & Le Bellac paper. Holland & Brown argued that the limit process applied to the scalar and vector potential would break gauge invariance such that they did not explore as Lévy-Leblond & Le Bellac the consequences for the so-called gauge conditions [3]. De Montigny & al. in a series of papers revisited also Galilean electromagnetism with the help of a "Galilean tensor calculus" which consists in expressing nonrelativistic equations in a covariant form with a five-dimensionnal Riemannian manifold ([4] and references therein). In their review on Galilean Electromagnetism, De Montigny & al. have shown that the Lorenz gauge condition $\nabla.\mathbf{A} + \frac{1}{c_L^2}\frac{\partial V}{\partial t} = 0$ which is covariant with respect to the Lorentz transformations becomes the Coulomb gauge condition $\nabla.\mathbf{A} = 0$ within the magnetic limit and that the Lorenz gauge condition keeps unchanged within the electric limit. The present author has reached independently the same conclusions [5] by imposing directly Galilean covariance with respect to the gauge conditions depending on the Galilean transformations for the potentials, first introduced by Levy-Leblond & Le Bellac, which differ according to the two limits :

<u>Electric Limit :</u>
$$V' = V \text{ and } \mathbf{A}' = \mathbf{A} - \frac{\mathbf{v}V}{c_L^2}$$

<u>Magnetic Limit :</u>
$$V' = V - \mathbf{v}.\mathbf{A} \text{ and } \mathbf{A}' = \mathbf{A}$$

by recalling that the Galilean transformations for the temporal and spatial derivations are :
$$\nabla = \nabla'$$
$$\partial_t + \mathbf{v}.\nabla = \partial_{t'}$$

Here we would like to show a physically meaningful derivation based on orders of magnitude of the Galilean limits for the Lorentz-covariant Lorenz gauge condition.

## 2. The Galilean limits of Lorenz gauge condition

Now, how do Lévy-Leblond and Le Bellac know that $E \ll c_L B$ or $E \gg c_L B$. Indeed, it is a rather formal assumption which is not justified at all a priori whereas it is true !

We argue that the derivation of Lévy-Leblond & LeBellac is equivalent to evaluate the order of magnitude of the following parameters :
$$\varepsilon = \frac{L}{c_L \tau} \text{ and } \xi = \frac{\tilde{j}}{\tilde{\rho} c_L}$$

where L ($\tau$ represents the order of magnitude of a typical scale (time) of the problem and $\tilde{j}$ ($\tilde{\rho}$) represents the order of magnitude of the current (charges) density in the system under examination.

As a matter of fact, the values of the electric and magnetic fields depend on their sources, that is, on the distribution of the charge and current densities. If one evaluates the order of magnitude of the fields in function of the sources using Gauss and Ampère's equations, one ends up with :
$$\frac{\tilde{B}}{L} \approx \mu_0 \tilde{j} \text{ and } \frac{\tilde{E}}{L} \approx \frac{\tilde{\rho}}{\varepsilon_0}$$

which leads to :
$$\frac{c_L \tilde{B}}{\tilde{E}} \approx \frac{\tilde{j}}{\tilde{\rho} c_L} = \xi$$

Hence, one has shown that assuming $E \gg c_L B$ ($E \ll c_L B$) is the consequence of assuming $\xi \ll 1$ ($\xi \gg 1$).

In addition, Ampère's equation leads to :
$$\tilde{B} \approx \frac{\tilde{v}\tilde{E}}{c_L^2}$$

and Faraday's equation gives :
$$\tilde{E} \approx \tilde{v}\tilde{B}$$

which are compatible only if : $\tilde{v} \approx c_L$ (Lorentz-covariant electromagnetism).

As a consequence, either we have $\tilde{B} \approx \frac{\tilde{v}\tilde{E}}{c_L^2}$ which is compatible with $\nabla \times \mathbf{E} \approx \mathbf{0}$ (the time derivative of the magnetic field drops) and $\nabla \times \mathbf{B} = \mu_0 \mathbf{j} + \frac{1}{c_L^2}\partial_t \mathbf{E}$ (the electric limit) or we have $\tilde{E} \approx \tilde{v}\tilde{B}$ which is compatible with $\nabla \times \mathbf{B} \approx \mu_0 \mathbf{j}$ (the time derivative of the electric field drops) and $\partial_t \mathbf{B} = -\nabla \times \mathbf{E}$ (the magnetic limit).

Once again, we underline forcefully that we have only shown compatibility between some approximations of the full set of "Maxwell equations" with Galilean relativity. We will now present what we think to be a demonstration of the two Galilean limits.

Indeed, the author has recently proposed to use the so-called Riemann-Lorenz formulation (the potentials are the basic quantities) instead of the so-called Heaviside-Hertz formulation (the fields are the basic quantities) in order to describe any experimental fact relative to Classical Electromagnetism [5]. The Riemann-Lorenz procedure consists in using the following postulate : "Any experimental fact of Classical Electromagnetism can be explained through the use of a scalar and a vector potential which are solutions of a set of Riemann equations with source terms (current density for the vector potential and charge density for the scalar potential) assuming that both potentials are constrained to fulfill the Lorenz equation. With respect to the interaction with the matter, the Lorentz force usually written in terms of the fields can be rewritten in terms of the potentials as the time derivative of the sum of the kinetic momentum plus the electromagnetic momentum equal to minus the gradient of the difference between the scalar potential and the scalar product of the velocity of the charge with the vector potential" :

$$\nabla^2 V - \frac{1}{c_L^2}\frac{\partial^2 V}{\partial t^2} = -\frac{\rho}{\varepsilon_0} \quad \text{and} \quad \nabla^2 \mathbf{A} - \frac{1}{c_L^2}\frac{\partial^2 \mathbf{A}}{\partial t^2} = -\mu_0 \mathbf{j} \text{ : Riemann equations}$$

$$\nabla \cdot \mathbf{A} + \frac{1}{c_L^2}\frac{\partial V}{\partial t} = 0 \text{ : Lorenz equation}$$

$$\frac{d}{dt}(m\mathbf{v} + q\mathbf{A}) = -\nabla(V - \mathbf{v}\mathbf{A}) \text{ : Lorentz force}$$

The purpose of this article is not to discuss the validity of this postulate but to show what it implies with respect to Galilean Electromagnetism using the potentials.

Assuming that the sources vanish at infinity, the potential are expressed by the so-called retarded formula :

$$V(M,t) = \frac{1}{4\pi\varepsilon_0} \iiint \frac{\rho(P, t - PM/c_L)}{PM} d\tau \text{ and } \mathbf{A}(M,t) = \frac{\mu_0}{4\pi} \iiint \frac{\mathbf{j}(P, t - PM/c_L)}{PM} d\tau$$

We explicitly assume that the potentials are defined up to a constant which, for an infinite volume, is taken to be zero. If the volume of investigation is bounded like in a Faraday cage, the contribution of all the sources outside the volume resumes to a constant which is different from zero as can be shown easily with the Green formula.

In the quasi-static approximation where $\varepsilon \ll 1$, the so-called retarded formula for the potentials become [6] :

$$V(M,t) \approx \frac{1}{4\pi\varepsilon_0} \iiint \frac{\rho(P,t)}{PM} d\tau \text{ and } \mathbf{A}(M,t) \approx \frac{\mu_0}{4\pi} \iiint \frac{\mathbf{j}(P,t)}{PM} d\tau$$

These approximations are the solutions of Poisson equations for the potentials which are the quasi-static limits of the Riemann equations with source terms [6] :

$$\nabla^2 \mathbf{A} \approx -\mu_0 \mathbf{j} \text{ and } \nabla^2 V \approx -\frac{\rho}{\varepsilon_0}$$

From this last remark, we can evaluate the order of magnitude of the potentials in function of the sources $\tilde{j}$ and $\tilde{\rho}$ which are given a priori :

$$\tilde{A} \approx \frac{\mu_0}{4\pi} \frac{\tilde{j} \vartheta}{L} \text{ and } \tilde{V} \approx \frac{1}{4\pi\varepsilon_0} \frac{\tilde{\rho} \vartheta}{L}$$

Contrary to Holland & Brown [3], we explicitly break gauge invariance of the Heaviside-Hertz formulation by giving orders of magnitude to the potentials. When one say that we can evaluate the order of magnitude of the potentials, one assume that we evaluate the order of magnitude of the potentials with respect to the constant on the boundary of the domain which is null if infinite and without sources at infinity. Hence, the tilde means the order of magnitude of a difference of potentials. Indeed, only the concept of difference of potential does have a physical meaning in the Riemann-Lorenz formulation. Yet, we point out forcefully that a difference of potential is not equal to a field : for example, the static field inside a capacitor is equal to the difference of potential between the two plates divided by the distance between them. A volt per meter is not the same object as a volt…

Now, one can form the following non-dimensional ratio :

$$\frac{c_L \tilde{A}}{\tilde{V}} \approx \frac{c_L \mu_0 \tilde{j}}{\frac{\tilde{\rho}}{\varepsilon_0}} = \frac{\tilde{j}}{\tilde{\rho} c_L} = \xi$$

We would like to know what become the Lorenz gauge condition as well as the charge conservation $\nabla \cdot \mathbf{j} + \frac{\partial \rho}{\partial t} = 0$ within the Galilean limits. We evaluate the orders of magnitude of each component of the spatial terms in these equations with respect to the temporal term :

$$\frac{\|\nabla \cdot \mathbf{A}\|}{\left\|\frac{1}{c_L^2} \frac{\partial V}{\partial t}\right\|} \approx \frac{\frac{\tilde{A}}{L}}{\frac{\tilde{V}}{c_L^2 \tau}} \approx \frac{c_L \tau}{L} \frac{c_L \tilde{A}}{\tilde{V}} = \frac{\xi}{\varepsilon} \text{ and } \frac{\|\nabla \cdot \mathbf{j}\|}{\left\|\frac{\partial \rho}{\partial t}\right\|} \approx \frac{\frac{\tilde{j}}{L}}{\frac{\tilde{\rho}}{\tau}} \approx \frac{c_L \tau}{L} \frac{\tilde{j}}{c_L \tilde{\rho}} = \frac{\xi}{\varepsilon}$$

As one can see, it implies the same ratio between $\varepsilon$ and $\xi$. Now, according to Lévy-Leblond & Le Bellac the quadri-current has the following Galilean limits :

Electric Limit :
$$\rho'=\rho \text{ and } \mathbf{j'}=\mathbf{j}-\rho\mathbf{v}$$
which leads to $\xi_e = \dfrac{\tilde{j}}{\tilde{\rho}c_L} \approx \dfrac{\tilde{\rho}\tilde{v}}{\tilde{\rho}c_L} \approx \varepsilon$

Magnetic Limit :
$$\rho'=\rho-\dfrac{\mathbf{v}\cdot\mathbf{j}}{c_L^2} \text{ and } \mathbf{j'}=\mathbf{j}$$
which leads to $\xi_m = \dfrac{\tilde{j}}{\tilde{\rho}c_L} \approx \dfrac{\tilde{j}}{\dfrac{\tilde{v}\tilde{j}}{c_L^2}c_L} \approx \dfrac{1}{\varepsilon}$

Hence, $\xi$ is different whether one considers the electric or the magnetic limit.

For Lorentz covariant Electromagnetism, we have obviously $\varepsilon \approx O(1)$ and $\xi \approx O(1)$ which implies that the two terms in the Lorenz gauge are of the same order of magnitude : Lorenz gauge condition is Lorentz covariant which is well known.

In the quasi-static approximation where $\varepsilon \ll 1$, we get :

Electric Limit :
$$\xi_e \approx \varepsilon \ll 1 \text{ and } \dfrac{\xi_e}{\varepsilon} \approx O(1)$$

The Lorenz gauge $\nabla\cdot\mathbf{A}+\dfrac{1}{c_L^2}\dfrac{\partial V}{\partial t}=0$ is now Galilean covariant with respect to the electric transformations of the potentials.

Magnetic Limit :
$$\xi_m \approx \dfrac{1}{\varepsilon} \gg 1 \text{ and } \dfrac{\xi_m}{\varepsilon} \gg 1$$

Hence, the Coulomb gauge $\nabla\cdot\mathbf{A}=0$ is the approximation of the Lorenz gauge within the magnetic limit and is now Galilean covariant with respect to the magnetic transformations of the potentials. The same conclusion applies for the charge conservation.

Using the Poisson equations for the potentials and either the Lorenz or the Coulomb gauge depending on the electric or the magnetic limit, one can easily derive the two sets of Galilean Maxwell equation for the fields proposed by Lévy-Leblond & Le Bellac [1]. The important point is to recognize that the two Galilean sets of equations in terms of the fields were stated without demonstration in [1] whereas here, we can demonstrate them starting with the potentials.

### 3. Conclusion

As a conclusion, we have shown that the Lorenz equation applies in both Lorentz-covariant relativity as well as Galilean covariant electric limit of Lévy-Leblond and Le Bellac whereas the Coulomb gauge equation applies only within the Galilean covariant magnetic limit. We have explicitly broken gauge invariance in order to get these results in accordance with the Riemann-Lorenz formulation of Classical Electromagnetism. This last fact is a priori astonishing and contradictory but it was demontrated long ago that Galilean Covariance and Gauge Invariance were incompatible [6]. Galilean Electromagnetism is an unexpected field of actual research as one needs to explore all its consequence in our current understanding of the special theory of relativity. As recalled recently by J. Norton, this theory emerged from Albert Einstein's struggle with Lorentz's pre-1905 electromagnetic theory, which is a mixing of the magnetic and electric limit without the essential property of group additivity and which made it untenable [7]…

**Bibliography**

[1] M. Le Bellac & J.-M. Lévy-Leblond, Galilean electromagnetism, Nuovo Cimento 14B, p. 217-233, 1973.

[2] J.R. Melcher & H.A. Haus, Electromagnetic fields and energy, Hypermedia Teaching Facility, M.I.T., 1998. Book available online at : http://web.mit.edu/6.013_book/www/ .

[3] P. Holland & H. Brown, The non-relativistic limits of the Maxwell and Dirac equations: the role of Galilean and gauge invariance, Studies In History and Philosophy of Science Part B: Studies In History and Philosophy of Modern Physics, Volume 34, Issue 2, p. 161-187, June 2003.

[4] M. De Montigny, F.C. Khanna & A.E. Santana, Nonrelativistic Wave Equations With Gauge Fields, International Journal of Theoretical Physics, Vol. 42, No. 4, p. 649-671, April 2003.

[5] G. Rousseaux, On the physical meaning of the gauge conditions of Classical Electromagnetism : the hydrodynamics analogue viewpoint, Annales de la Fondation Louis de Broglie, Volume 28, Numéro 2, p. 261-270, 2003. Article available online at : http://www.ensmp.fr/aflb/AFLB-282/ablb282p261.htm

[6] S.K. Wong, Gauge invariance and Galilean invariance, Nuovo Cimento 4B, p. 300-311, 1971.

[7] J.D. Norton, Einstein's Investigation of Galilean Covariant Electrodynamics Prior to 1905, Archive for History of Exact Sciences, Volume 59, Number 1, p. 45-105, November 2004.